\documentclass[reprint,aps,prb,amsmath,amssymb]{revtex4}

\usepackage[T1]{fontenc}
\usepackage[]{graphicx}
\usepackage{times}
\usepackage{bm}
\usepackage{color}

\newcommand{\comment}[1]{}

\newcommand{\be}{\begin{equation}}
\newcommand{\ee}{\end{equation}}
\newcommand{\bea}{\begin{eqnarray}}
\newcommand{\eea}{\end{eqnarray}}

\newcommand{\dittostraight}{---\textquotedbl---}

\begin{document}

\title{Exciton complexes in low dimensional transition metal dichalcogenides}

\author{A. Thilagam}
\email[]{thilaphys@gmail.com}

\affiliation{Information Technology, Engineering and Environment, 
University of South Australia, Australia
 5095}

\begin{abstract}
We examine the excitonic properties of layered configurations of low dimensional
transition metal dichalcogenides (LTMDCs) using the fractional dimensional
space approach. 
The binding energies of the exciton, trion and biexciton
in LTMDCs of varying layers are analyzed, and linked to the dimensionality
parameter $\alpha$, which provides insight into critical electro-optical properties
(relative oscillator strength, absorption  spectrum, exciton-exciton interaction)
of the material systems. The usefulness of $\alpha$  is highlighted by its
independence of the  physical mechanisms  underlying
the confinement effects of
 geometrical structures. Our estimates
of the binding energies of exciton complexes for the monolayer configuration
of transition metal dichalcogenides
 suggest a non-collinear structure for the trion  and
a positronium-molecule-like square structure
for the biexciton.
\end{abstract}

\maketitle

\section{Introduction}
Transition metal dichalcogenides 
 MX$_2$ (Transition metal M = Mo, W, Nb; Chalcogen X = S, Se), \cite{wilson,lieth,butler,pan2014}
 are currently  studied with renewed interest \citep{pan2014,gana} due to their
attractive opto-electronic properties, \cite{he2014stacking,wang,pu2014flexible,Chang2013,kumar2014tunable,johari2011tunable}
and advances in fabrication techniques  involving exfoliation into single and multilayer 
layered configurations known as  LTMDCs \cite{cho2008,beck2000,mouri2013,joensen1986single}.
These material systems present potential
applications in optics,  \cite{zhang2014m,lopez2014light,sobhani} sensing,  \cite{liu2012bio,wu2010graphene,li2014preparation}
chemical and biological systems, \cite{li2014preparation,ou2014ion}
 and quantum informatics and spintronics \cite{fogler2014,xu2014spin,glazov2014}.
In  TMDCs,  the M metal layer is sandwiched between two X layers
with strong intra-plane  bondings and
weak inter-plane interactions giving rise
to the low dimensional, highly anisotropic hexagonal arrangement of   M and X
atoms with spacegroup P6$_3$/mmc associated and the space group number
194. While TMDCs exhibit some properties
that complement those of  graphene which possess a similar honeycomb pattern
of arrangement of atoms, there exist other unique features that do not match.
For instance, there is   absence of bandgap in  graphene while
TMDCs are amenable to band-gap engineering properties, \cite{wang,mak,conley2013}
with potential to act as suitable inorganic substitutes
for graphene based applications.

Depending on the number of lattice layers constituting the material,
LTMDCs display  crossovers in the spectral  characteristic of their band-gap  \cite{splen,makatom,zhao}.
Molybdenum disulfide (MoS$_2$) which is  a well known prototype of the LTMDCs, 
appears as an  indirect band-gap semiconductor  in its
 bulk form, but transforms  to
a direct band-gap material as the material thickness is decreased
 due to the inversion symmetry breaking of the lattice structures \cite{makatom}.
The presence of direct band-gap is evidenced by strong photoluminescence 
in the vicinity of the $K$ and $K'$ points of the Brillouin zone \cite{splen,makatom}.
Similar crossovers are also observed
 in WSe$_2$ and WS$_2$ \cite{zhao} as the thickness of the material is reduced.

 Excitonic features  are pronounced in monolayer LTMDCs
as a result of their large  binding energies   which occur
due to two main effects: 
(i)  decreased  thickness of the material system results in
increased  interactions between charges and a dimensionality dependent
macroscopic surface self-polarization term \cite{delerue}
gives rise to an  increased electron-electron correlation
and reduced screening effects. The subsequent
increase in the quasiparticle gap can be correlated with high exciton
binding energies \cite{wirtz2006}.  (ii) The enhanced electron-hole
overlap due to the geometrical confinement of exciton wave functions
in the reduced dimensional space \cite{still,lefeb,thilpho} results in
large exciton binding energies \cite{he,mathieu}.
The existence of  robust
electron-hole correlated states in inorganic   LTMDCs provide
viable alternatives to stable exciton states which commonly occur
 in organic molecular systems. 
The geometrical confinement of exciton wave functions  within a small space
region  enhances optical densities and 
oscillator strengths \cite{christo93} and  supports the  existence of exciton
complexes such as trions \cite{mak,berkel13,wang2014non} and biexcitons \cite{siebiex} that 
are  stable at room  temperatures.

Using two-photon excitation
spectroscopy and  density functional methods based on the
 GW-Bethe Salpeter equation (GW-BSE) approach, 
 the exciton binding energy was estimated to be as high  as 0.7 eV \cite{ye2014probing}
in Tungsten Disulfide (WS$_2$).  In  Molybdenum Disulfide (MoS$_2$),
the incorporation of electron-hole interactions within
the BSE approach revealed  $A$ and $B$ excitons 
linked  to two pronounced peaks at respective 
positions, 1.78 eV and 1.96 eV \cite{rama}. These spectral features are
consistent with experimental
results, \cite{makatom}  and yield exciton binding energies  of 0.85 eV, \cite{rama}
in agreement with the binding values computed by Cheiwchanchamnangij et. al. \cite{chei12}.
First-principles calculations  based on the
GW-BSE approach, and  incorporating excitonic and  electron-phonon effects 
yield exciton binding energies  of 0.96  eV \cite{qiu2013}.
By including  both excitonic effects and spin-orbit coupling and larger
$k$-grid mesh which ensures good convergence during computations, much lower binding energies ($< $ 0.4 eV) 
were obtained 
\cite{molinaspin}, with $A$, $B$ exciton spectral positions appearing
to be in  good agreement with experimental results.
The incorporation of
charge carrier-light coupling element which incorporate optical selection  and polarization rules,
yield  binding energies in the range 0.42-0.44 eV  \cite{berg}.
It thus appear that   precise
estimates of the exciton binding energies 
in monolayer LTMDCs remain elusive, with a
 range of binding energies predicted  for the prototype MoS$_2$ of varying
dimensions. A full comparison of the binding energies from various groups
which utilized  first-principles many-body calculations and experimental
techniques are  shown in Table-I of Section \ref{frac}.

The discrepancies in the
exciton binding energies reported by different groups   may arise
from technical errors such as low $k$-point sampling \cite{rama} and neglect
of spin-orbit interaction as discussed in Ref.\cite{molinaspin}.
Exciton binding energies require a dense $k$-point mesh to converge and 
produce reliable results \cite{huser13diele}, which can be computationally demanding.
The   electronic processes which occur
on the surfaces of the slab-like structures of LTMDCs
differ from those in bulk solids, with a potential source of
 error due to inaccurate estimation of the macroscopic screening in systems with 
finite width. The shift of surface plasmons  due to confinement
effects \cite{kumar2012tunable} shows the need to account 
for convergence of the electronic
 surface  with respect to number of monolayers
 in many-body calculations. Currently available
density functionals  do not fully incorporate
the  strong Coulomb interactions and non-local correlations
 between the 3$d$ and 4$s$ electrons of the transition metal atoms.

In  Molybdenum Disulfide (MoS$_2$), the intra-plane  bonding  within the S-Mo-S structure is known to
give rise to  mixing of $s$, $p$ and $d$ orbitals of Mo
with the $p$ orbital of S \cite{fleis,mcmenamin}.  These mixing effects 
 change when the number of lattice layers making up the material system  is altered.
There are corresponding changes to the surface charge distribution, which are normally not fully incorporated in
density functional computations. Therefore, there exist
several sources of errors  that need to be addressed before  exciton binding energies
can  evaluated with greater accuracy
in  LTMDCs. Current experimental techniques are unable to perform a direct measurement 
of exciton binding energies, which underpins the importance of {\it ab initio} 
computational methods that provide an indirect mode of evaluating the binding energy.

In this study, we adopt an approach
based on fractional dimensions \cite{still} to examine
exciton binding energies in LTMDCs. The fractional dimensional space
approach (FDSA)  has  been utilized
in  studies of excitonic properties 
\cite{he,matos99,christo93,mathieu,lefeb,oh1999geometric,thilpho,reyes2000exci,thilexci,thil97stark} 
by fitting to  models that utilizes a variable dimension, $1 \le \alpha \le 4$
which provide good agreement with experiments findings. 
The pseudo-two-dimensionality attribute of the exciton \cite{he,lefeb}
simplifies the evaluation of electro-optical properties.
This is performed by  the mapping of the  anisotropic exciton 
 in the real space into an isotropic environment parameterized by a single quantity, $\alpha$.
The usefulness of $\alpha$ can be appreciated by the fact that
it is independent of the  physical mechanisms   governed
by  confinement effects arising from specific
geometrical structures. Consequently a particular optical 
 spectrum can be attributed to
different systems as long as these possess the same dimensionality.
In earlier works, the exciton binding energy and characteristics of the 
optical spectrum, \cite{he,matos99,christo93,mathieu,lefeb,mohapat2014}
 exciton-exciton interaction strength, \cite{thilexci} properties of donor states, \cite{mat01frac}
 and exciton-phonon coupling \cite{thilpho}  have been
 determined using the dimensionality parameter, $\alpha$.

The  parameter $\alpha$ takes into account the 
tunneling effect \cite{thilexci} so that the  exact
two-dimensional exciton model is best valid  at a
critical confinement length, depending on the material composition and other external conditions
related to electric and magnetic fields.
The use of a rigid exciton model for which $\alpha$ = 2 or 3  can give rise to an
overestimation of the exciton binding energies.
Excitons described by fractional values of $\alpha$ are able to provide realistic
values of the binding energy, however it is important 
 that $\alpha$  is defined using
appropriate material parameters. 
In this study, we show that the FDSA provides an intuitive approach to understanding the
optical features that are affected by a change in the number of layers
comprising LTMDCs. We highlight the convenient estimation of the binding energies
 of charged exciton complexes \cite{thiltrion}
and biexcitons \cite{ron12bi,oh1999geometric,birk1996bind}
using  analytical relations based on the FDSA.

\section{Excitons in fractional dimensional space}\label{frac}

In a fractional dimensional space described by
the parameter $\alpha$, the relative motion
of a correlated electron-hole pair with generalized
 coordinates $r, \theta$ appear as \cite{still,lefeb}
\bea 
\left[ -\frac{\hbar^2}{2 \mu} 
\frac{1}{r^{\alpha-1}} \frac{d}{d r} 
\left(r^{\alpha-1} \frac{d}{d r} \right) - \frac{e^2}{4 \pi \epsilon r}
-\frac{\hbar^2}{2 \mu r^2}\frac{1}{\sin^{\alpha-2} \theta}. \frac{d}{d \theta}
\left(\sin^{\alpha-2} \theta \frac{d}{d \theta} \right) \right ] && \\ \times \Phi(r,\theta)
= E. \Phi(r,\theta),
\label{fraeq}
\eea
where the exciton reduced mass  $\mu^{-1} = m_e^{-1}+m_h^{-1}$, $m_{e}$ and $m_{h}$  being the 
effective masses of the electron and hole, respectively. The exciton wavefunction can be separated
as  $\Phi_n(r,\theta) = \psi_{n}(r) \Psi(\theta)$ where $\Psi(\theta)$ denotes an eigenfunction of the angular
momentum $L^2$, $L^2 \Psi(\theta) = l(l+\alpha-2) \Psi(\theta)$ \cite{lefeb}. 
The quantum number, $n = 1,2,\cdots$, is the 
principal quantum number state. The decoupling of $\Phi_n(r,\theta)$
allows Eq. \ref{fraeq} to be split into two independent equations \cite{still,lohe2004},
with only one that depends on the energy of the exciton.
The 1s state of an exciton in an $\alpha$D space appear as
\be
\label{wave1}
\psi_{1s}(r) = \left [ {2 ^ {\alpha+1} \; \pi^{{1-\alpha \over2}} 
\over   \Gamma[{\alpha-1 \over 2}] \; (\alpha - 1)^{\alpha+1}} \;
\;  {1 \over a_{_B}^\alpha } \right ]^{1 \over 2} \; \exp \left[ - {2 \over {\alpha-1}}\; 
{r \over a_{_B}} \right ] 
\ee
with the  binding energy \cite{still,he}
\be
E_{bX}^c = {R_y \over \left (n+{\alpha-3 \over 2} \right )^2},
\label{be}
\ee
where $a_{_B}$ = 0.529  $\frac{\epsilon}{\mu}$ \AA  \; is the three-dimensional Bohr radius 
of the exciton,  and $R_y$ = 13.6 $\frac{\mu}{\epsilon^2}$ eV is the effective exciton Rydberg energy.
The exciton reduced mass  $\mu$ is given in terms of the free-electron mass.
The parameter  $\alpha$ is dependent on the variables  used to describe the 
dimensionality of the confined exciton. The  mapping of layered low dimensional LTMDCs into
 the effective fractional-dimensional  space may be performed
by  utilizing the  dimensionality  based on the ansatz \cite{christo93,mat02pol,mat01frac}
$ \alpha = \gamma_x + \gamma_y +\gamma_z $
where $\gamma_i$ (i = x,y,x) denotes  the measure of  confinement
in the $x$, $y$, and $z$, respectively.  As 
there is no restriction to motion in the $x,y$ plane of layered structures, we 
get $\gamma_x =\gamma_y = 1$. In the direction where there is 
confinement, we employ the relation, \cite{christo93,mat02pol}
$  \gamma_z = 1-\exp[-\chi]$ where
\be
  \chi = \frac{\textrm{confinement length}}{\textrm{effective  interaction
length }}
\label{chi}
\ee
The confinement of excitons to lower dimensions $\alpha$ = $3-\exp[-\chi]$ can be seen
as increasing the  electron and hole overlap wavefunctions in $\exp[-\chi]$
dimensions, which may assume fractional values. This gives rise to an enhanced 
Coulomb interactions between the charge carriers resulting in increased exciton
binding energies. 
Using Eq. \ref{chi}, we obtain the well known expression for $\alpha$ applicable to
 two-dimensional heterostructures such as the infinite quantum
wells of confinement width $L_w$  \cite{still,lefeb}
\be
\alpha = 3 - \exp(-{L_w \over 2 a_{_B}}).
\label{alpha}
\ee   
Eq. \ref{alpha} needs to be  modified in the case of finite quantum wells  where there is 
penetration of the carriers wavefunction 
into regions outside the well. 
In LTMDCs where there is  crossover from an indirect band-gap to a 
a direct band-gap with decreased thickness,  \cite{splen,makatom,zhao} 
 carrier dependent parameters such as the effective mass, dielectric constants, 
and location of $K$ points need  to be incorporated for accurate evaluation of $\alpha$. In this study, we propose
the dimensionality, $\alpha$, based on the ratio of the effective exciton Rydbergs, $\beta$
\bea
\label{alphaR}
\alpha \approx 2 \sqrt{\beta}+1 \\
\beta = \frac{R_y^{3D}}{R_y^{\alpha D}}
\label{Ryd}
\eea
In bulk material systems, $R_y^{\alpha D} \rightarrow R_y^{3D}$,  we obtain $\alpha$ = 3
and  $E_{bX}^c$ = $R_y^{3D}$ for the ground state ($n$ = 1). 
In lower dimensional systems such as monolayer and bilayer LTMDCs,
the  dielectric constant is decreased resulting in larger effective Rydbergs, $R_y^{\alpha D} > R_y^{3D}$,
with  $\alpha < 3$.  Eq. \ref{alphaR} incorporates the   dependency of  $\alpha$ on the 
dielectric environment and effective exciton mass
altered by  confinement effects. 

In Table-I, we  compare the  exciton binding energies at the 
$K$ high symmetry points of the Brillouin zone of monolayer MoS$_2$ reported in
earlier studies, and our estimates obtained by  using available $\mu, \epsilon$ and
$E_{bX}^c$  = 13.6 $\mu/\epsilon^2$ eV. We  provide brief descriptions
related to the convergence parameters, as well justification for
our use of the dielectric parameters in Table-I. 
The  exciton radius $\approx$ 9 \AA \; appears to be a consistent 
estimate in earlier works as well as from our choice of
 $\mu, \epsilon$  for the monolayer MoS$_2$ system.
The change in dimensionality $\alpha$ of materials
of varying layer numbers is analyzed using Eqs.\ref{alphaR} and \ref{Ryd}.
The dimensionality, $\alpha$ = 3  is computed using Eqs.\ref{alphaR} and \ref{Ryd}, with
the bulk Rydberg $R_y^{3D}$  estimated using
$\mu$ = 0.4 \cite{fortin75} and $\epsilon$ = 10.71 \cite{chei12}.
We expect slight variations in $\alpha$ to occur for
$\mu$, $\epsilon$  that depart from our estimates used to 
compute $R_y^{3D}$. Nevertheless, the qualitative features
highlighted by the increase in $\alpha$ with increase in the 
number of layers is expected to remain intact.
The main result of this study is thus the demonstration of the
gradual increase in $\alpha$ from 2 in monolayers, to intermediate
values (2.30 < $\alpha$ < 2.69)  in bilayer and 4-layer  MoS$_2$ system to the standard reference,
$\alpha$ = 3 in bulk  MoS$_2$. It is to be noted that we 
have employed $\mu$ in the range (0.2 - 0.3) to compute $\alpha$
for the bilayer, 4-layer and 6-layer MoS$_2$ system, due to unavailability
of $\mu$ values for these systems.

Table-I highlight the  discrepancies between the  binding energy estimates in some earlier works
and our estimates based on $E_{bX}^c$ for the monolayer MoS$_2$ system.
By using $\alpha$ = 3 in bulk  MoS$_2$,
the energy range of 0.2 - 0.3 eV ($\alpha \approx$ 2) provides   reasonable estimates
of the binding energy of MoS$_2$ in the monolayer configuration at $\alpha$ = 2. 
Hence one underlying reason for  disparities
in binding energies for the monolayer configuration as shown in 
Table-I can be attributed to an underestimation of the
 dimensionality parameter $\alpha$ in the MoS$_2$ system. 
This gives rise to an overestimation of the exciton binding energy
as noted in some works. Other contributing factors
may arise  due to variations in experimental conditions, convergence criteria, dielectric
tensor estimates, choice of the crystal  lattice constants (a and c) and 
the separations between  the monolayers.
\begin{table}
    \caption{\label{tab:gw}
Comparison of exciton binding energies at the 
$K$ high symmetry points of the Brillouin zone of monolayer MoS$_2$ based 
on earlier studies and our theoretical prediction using $E_{bX}^c$ 
= 13.6 $\mu/\epsilon^2$ eV. The change in the dimensionality $\alpha$ of materials
of varying layer numbers is analyzed using Eqs.\ref{alphaR} and \ref{Ryd}.
 The exciton effective mass is denoted by $\mu$ and the ratio $\sigma$ = $\frac{m_e}{m_h}$.
The exciton effective mass and dielectric constants  employed to determine the energy estimates
are included. In Ref.\cite{molinaspin}, the density-functional theoretical calculations
included spin-orbit coupling effects, using a large $k$-grid mesh ($>$ 18 $\times$ 18).
The GW-Bethe Salpeter equation (GW-BSE) approach in Ref.\cite{qiu2013} incorporates  self-energy, excitonic,
and electron-phonon effects but utilizes comparatively smaller $k$-grid mesh (12 $\times$ 12).
Refs.\cite{chei12,rama} adopt small   $k$-grid mesh ($\le$ 12 $\times$ 12)
while the computation in Ref.\cite{shi2013quasi} involve a  $k$-grid mesh of  15 $\times$ 15.
The electron and hole masses provided for Ref.\cite{chei12} are the average
of the $K_l$ (longitudinal) and $K_t$ (transverse)  symmetry points of the Brillouin zone.
The pair of exciton mass values for Ref.\cite{berg} correspond respectively to the renormalized
spin-down and spin-up valence effective band masses due to spin-orbit coupling,
and the pair of binding energies correspond respectively to those of the $A$ and $B$ excitons.\cite{berg}
 \newline
 The effective
dielectric constant, $\epsilon$ = $\sqrt{\epsilon_\bot \; \epsilon_\|}$ where $\epsilon_\bot$ ($\epsilon_\|$)
denote dielectric  components perpendicular (parallel) to the lattice vector in the $z$ direction.
$E_{bX}$ is  the exciton binding energy 
that appear in the cited references, which is compared with our  estimate of the binding energy $E_{bX}^c$ 
= 13.6 $\mu/\epsilon^2$ eV  using $\mu$ and  $\epsilon$ provided in the references.
Additional notes are provided below when $\mu$ and  $\epsilon$ cannot be retrieved from the references.
Physical quantities appearing with superscript $*$ are our  estimates of $\mu$, $\epsilon$ used
to complete evaluation of   $E_{bX}^c$.
The dimensionality, $\alpha$ is computed using Eqs.\ref{alphaR} and \ref{Ryd}, with
the bulk Rydberg $R_y^{3D}$ ($\alpha$ = 3) based on
$\mu$ = 0.4 \cite{fortin75} and $\epsilon$ = 10.71 \cite{chei12}.
The  trion binding energies are estimated using Eq. \ref{tri2}.
All binding energies  appear in units of  eV.
\newline
$^\diamond$ The experimental value of 0.08 eV  in bulk  MoS$_2$ is provided in Ref.\cite{komsa2012effects}\newline
$\ddagger$ \; $\epsilon_\bot$ = 3.76 is obtained using Eq.(30) of  Ref.\cite{qiu2013}.\newline
$\dagger$ \; The binding energy of 0.96 eV in Ref.\cite{qiu2013} corresponds to a $k$-grid mesh of  72 $\times$ 72
while the effective mass parameters are determined on a 12 $\times$ 12 $k$-grid. \newline
$\ddagger \ddagger$ We use $\epsilon$ = 3.43 \cite{chei12}  (not specified in  Qiu et. al. \cite{qiu2013}) and $\mu$ = 0.2 to compute
 $E_{bX}^c$ (Eq. \ref{be}) and   $E_{bX^-}$ (Eq. \ref{tri2}). 
The  exciton radius = 9 \AA \; obtained
using these values of $\mu, \epsilon$  is consistent with the 
root mean square exciton radius of 10 \AA \; by Qiu et. al. \cite{qiu2013}.\newline
{}$^a$ $\mu, \sigma$ are computed using  band edge masses, $m_e$ = 0.5, $m_h$ = 0.6 retrieved from Zhang et. al. \cite{zhang2014}. \newline
{}$^b$ We determine $\epsilon$ = 3.68  (not specified in  Zhang et. al. \cite{zhang2014}) based on  the measured exciton radius = 9.3 \AA \; at electron density $\approx$ 2-4 $\times$ 10$^{12}$/cm$^2$
in Ref.\cite{zhang2014}. 
\newline
{}$^c$ We determine $\alpha$ and $E_{bX}^c$ based on $\mu$ = 0.27 {}$^a$ and $\epsilon$ = 3.68 {}$^b$.
\newline
{}$^{d3}$ We compute $\alpha$ and $E_{bX}^c$ based on $\mu$ = 0.28$^{d1}$ and 
the transverse component of the macroscopic static dielectric tensor, $\epsilon$ = 3.68$^{d2}$.
\newline
{}$^e$ $\mu, \sigma$ are computed using electron and hole effective masses 
at $K$ points for optimized lattices,  $m_e$ = 0.32, $m_h$ = 0.37 obtained from Shi et. al.\cite{shi2013quasi}
\newline}
\begin{tabular}{|c|c|c|c|c|c|c|c|c|c|c|c}
\hline       \hline
		 ~ & No. of Layers &$\mu$ & $\sigma$& $\epsilon_\bot,\epsilon_\|,\epsilon$  & $E_{bX}$& $\alpha$
(Eqs.\ref{alphaR}, \ref{Ryd})
\newline
 & $E_{bX}^c$ (13.6 $\mu/\epsilon^2$) &  $E_{bX^-}$ (Eq. \ref{tri2})\\ 
        \hline 
          Qiu et. al. \cite{qiu2013} & monolayer & 0.13 & -- &  3.43$^{\ddagger \ddagger}$ & 0.96$^\dagger$ &  1.95 & 0.21 & 0.021 \\ 
Cheiwchanchamnangij et. al.\cite{chei12} & monolayer & 0.19 & 0.80 & 4.2, 2.8, 3.43 & 0.897 & 1.92 & 0.22 & 0.022 \\
\dittostraight &bilayer & 0.21 & 0.82 & 6.5, 4.2, 5.22 & 0.424 & 2.34 & 0.11 & 0.011 \\
\dittostraight & bulk &  0.4 \cite{fortin75} & 0.80 & 13.5, 8.5, 10.71& 0.025 & 3.0 & 0.04 (0.08$^\diamond$) & < 0.005  \\
Berkelbach et. al.\cite{berkel13} & monolayer & 0.25 & -- & 3.76$^\ddagger$ & 0.54  &1.89 & 0.24 & 0.024 \\
& & & & &  & &  &
 (0.026 - 0.032) \cite{berkel13}\\
      Molina-S{\'a}nchez et. al. \cite{molinaspin,mol11phonons} & monolayer & 0.2$^*$ & -- & 7.36, 1.63, 3.46  & $<$ 0.4 & 1.91 & 0.23& 0.023  \\
 Zhang et. al. \cite{zhang2014} & monolayer & 0.27$^a$ & 0.83$^a$ &3.68$^b$ & 0.28 - 0.33 & 1.83$^c$  &0.27$^c$  & 0.03 \cite{zhang2014} \\
       Bergh{\"a}user \cite{berg} & monolayer & 0.28,0.26  & 0.73, 0.83 & -- & 0.42, \;0.44 & -- &-- &-- \\
        Ramasubramaniam \cite{rama} & monolayer & 0.28$^{d1}$ & 1.11 & 4.26$^{d2}$ & 0.85 & 1.95$^{d3}$  & 0.21$^{d3}$ & 0.021 \\
       Shi et. al.\cite{shi2013quasi} &  monolayer & 0.17$^e$ &  0.86$^b$  & -- & 0.54 & -- &-- &--  \\
       Kumar et. al. \cite{kumar2012tunable}  &  bilayer & 0.20 -  0.25$^*$ &  $\sim$ 0.8$^*$  &  6.9, 4.4, 5.51 & -- & 2.45 - 2.30& 0.09 - 0.11 & < 0.01 \\
\dittostraight  &  4-layer & {0.25 - 0.3}$^{*}$ &  $\sim$ 0.8$^*$  &  8.7, 5.9, 7.16 &--& 2.69 - 2.42& 0.07 - 0.09&  < 0.01 \\
\dittostraight &  6-layer & 0.25 - 0.3$^*$ &  $\sim$ 0.8$^*$ &  9.8, 6.4, 7.92 &--& 2.87 - 2.58& 0.05 - 0.08 & < 0.01 \\
Mak et. al. \cite{mak}, Singh et. al. \cite{singh2014coherent} & monolayer & -- & -- & -- &  --& -- &-- & 0.020 \cite{mak}, 
0.030 \cite{singh2014coherent} \\
        \hline
        \hline
 \end{tabular}
 \end{table}

\subsection{Change of dielectric properties with $\alpha$}

Table-I shows the difference in dielectric constants between  the monolayer,  
bilayer and the bulk MoS$_2$ system. There is decrease 
in dielectric components, $\epsilon_\bot,\epsilon_\|$
 with the dimensionality $\alpha$, which is consistent in all cited works.
In low-dimensional LTMDCs, the decrease of $\epsilon$ with $\alpha$ 
can be examined by considering each monolayer 
 of  M (= Mo, W, Nb) and  Chalcogen X (= S, Se)
atoms as a   homogeneous dielectric slab 
 of thickness $t$ with a bulk dielectric constant $\epsilon_m$. We note that in practical situations, the 
monolayer LTMDC material is sandwiched between 
two regions of different  dielectric constants.
For the simple model configuration of repetitive multi-layers with 
a vacuum interlayer separation, $D$,
the composite dielectric
components for the configuration of repetitive layers appear
as  $\epsilon_\|$ = $\eta  \left(\epsilon _m-1\right)+1$,
$\epsilon_\bot^{-1}$ = $1-\frac{\eta  \left(\epsilon _m-1\right)}{\epsilon _m}$ \cite{freyso}
 where $\epsilon_\bot$ ($\epsilon_\|$)
denote dielectric  components parallel (perpendicular) to the layers.
The parameter $\eta$ = $\frac{t}{t+D}$  provides the anisotropic measure of 
the layered system, so that $\eta$ = 1 represents the bulk system 
and a critical $\eta=\eta_c$ corresponds to  the monolayer configuration.
By setting $\epsilon_m$ = 10.71 \cite{chei12} and evaluating the relations
$\epsilon_\|$, $\epsilon_\bot$ as function of $\eta$,
the  interpolation between the bulk  MoS$_2$
and the monolayer LTMDC can be analyzed to reveal
the decrease of $\epsilon$ with $\eta$ (or $\alpha$).

\subsection{Change of absorption coefficient with $\alpha$}

The excitonic photoluminescence spectra can be analyzed via
the absorption coefficient derived in fractional dimensional space \cite{lefeb}
\bea
\nonumber
O(\hbar \omega) &=& O_o \left[\sum_{n=1}^\infty \frac{F_1(n, \alpha) R_y}
{ \left (n+{\alpha-3 \over 2} \right )^{\alpha+1}} \delta(\hbar \omega - E_n)
+ |\Gamma(\frac{\alpha-1}{2} + i \gamma)|^2 \; \frac{e^{\pi \gamma}}{\pi} \; \gamma^{2-\alpha} \;
\Theta(\hbar \omega)
\right ] \\ 
\label{absorp} \\
\label{absorp1}
O_o &=& \frac{F_2({\alpha}) \; |d_{cv}|^2 L_c^{\alpha-2}}{n_{_B} c m_o^2 \omega a_{ex}^\alpha 
R_y}
\eea  
where $\hbar \omega$ is the photon energy, and
$F_1(n, \alpha)$ and $F_2(\alpha)$ are explicit functions of $\alpha$.
$\Theta(\hbar \omega)$ is the Heavyside step function, the reduced parameter $\gamma = \sqrt{R_y/\hbar \omega}$
and $n_{_B}$ is the refractive index of the material. $L_c$ is the effective length of the
material, $|d_{cv}|^2$ is the conduction-to-valence squared matrix
element of the electron dipole moment,  $c$ is the velocity of light,
and $\Gamma(x)$ is the Euler gamma function. The absorption coefficient in
Eq. \ref{absorp} reduces to the expected expression in the exact 2D and 3D cases.

The parameter $\alpha$ establishes a 
strong correlation  between the exciton binding energy and the
the absorption coefficient in Eq. \ref{absorp}.
According to the $\alpha$ values evaluated for the number of layers
making up the MoS$_2$ system in Table-I, it can be seen
that there is  decrease in  $\alpha$ from the bulk, to the 4-layer
and bilayer and finally the monolayer configuration.
Following Eq. \ref{absorp} , we expect to obtain an enhancement in the 
 optical densities and oscillator strengths as the number of
layers is decreased until the 
monolayer configuration is reached. Likewise, changes
in exciton-phonon interactions \cite{thilpho},
stark-shifts due to the application of an external electric field \cite{thil97stark}
and exciton-exciton interaction strength, \cite{thilexci}
are expected when the number of layers in LTMDCs are reduced.

\section{Trions and biexcitons in LTMDCs}

\subsection{Charged excitons}
Charged exciton complexes such as negative $X^{-1}$ and positive $X^{+1}$ trions 
are formed when a single exciton is correlated respectively,
with  an additional  electron or hole.  Two-dimensional
 trions have a binding energy which is about
an order-of-magnitude larger than that of the bulk trions 
due to confinement effects. In a recent experiment  involving 
monolayer MoS$_2$ \cite{mak} there was
formation of stable trion  states with a binding energy of 20 meV
when the population of excess electrons was increased.
A higher binding energy of 30 meV was noted
in a recent  work  \cite{singh2014coherent} showing
distinct exciton and trion spectral
resonances. It would be worthwhile to compare these
results with theoretical models of the trion, with focus
on the negative trion consisting
of one hole that is shared by  two electrons.

In our earlier work, the collinear structure
of the trion was used to derive 
as simple relation relating the ratio of the trion to exciton binding
energies  as  \cite{thiltrion}
\be
\label{tris}
\frac{E_{bX^-}}{E_{bX}}=
\frac{9}{4} \left(\frac{2 \sigma +1}{\sigma ^2+4 \sigma +2}+1\right)^{-1}-1
\ee
Eq. \ref{tris} is applicable to the 
trion configuration with the
hole (electron) at the midpoint of two electrons (holes)
in negatively (positively) charged trion. 
In the case of the monolayer MoS$_2$, $\sigma \approx$ 0.8 - 1,
and we obtain typical trion binding energies, $E_{bX^-} \approx$
0.5 $E_{bX}$, which compares well with $E_{bX^-} \approx$
0.4 $E_{bX}$  obtained using the variational quantum Monte Carlo approach \cite{ronnow}.
These relations would however yield  much higher binding energies $\sim 100$ meV
(than currently observed 20 to 30 meV \cite{mak,singh2014coherent})
for typical exciton binding energy of $>$ 0.2 eV in the
monolayer MoS$_2$ system.  Therefore the collinear structure of the trion does
not adequately describe the charged exciton complex in LTMDCs.

Due to  the Pauli exclusion principle,  the multiple occupation
of single fermion  states is forbidden. The trion  trial wavefunction   
should therefore
incorporate parameters linked to
the phase space surrounding the exciton \cite{thilpauli,thiljmc}
for accurate energy convergence in variational calculations \cite{ronnow}. 
To this end, the excess electron added   to form a
trion is more likely to move in the direction perpendicular to the monolayer
surface,  and  bind to the  two-dimensional exciton located  within the monolayer.
A hyper-spherical approach  considered by Ruan et. al. \cite{ruan1999}
incorporates   the possibility of
modification of charge  interactions in the $xy$ plane due to
existence of charges outside the monolayer. Ref.\cite{ruan1999} predicts
the following relation between the trion and exciton binding energies
\be 
\label{tri2}
E_{bX^-} \approx 0.1 E_{bX} \quad \quad  \sigma > 0.6
 \ee
Eq. \ref{tri2} 
yields trion binding energies of  20-30 meV
which are consistent with experimental
observations \cite{mak,singh2014coherent}.
In Table-I, we have compared the 
negative charged exciton binding energies (Eq. \ref{tri2}) with experimentally observed values,
and also provided estimates of the binding energy
for other configurations of the  MoS$_2$ system.
The negatively charged trion may be difficult to resolve in the
high dimensional ($\alpha$ > 2.5) bulk and systems which exceed  4-layer thickness,
however there is good
possibility of its detection in 2-layer materials.
The predicted trion binding energies for other systems are presented 
in Table-II.  Similar to the  MoS$_2$ monolayer system,
the mass ratio $\sigma$ is close to  unity for other 
LTMDCs. This fact will be employed in the evaluation of 
the binding energies of  the biexciton in the next section.

\subsection{Biexcitons}

The energy of the biexciton,  a four-body system of 
 two electrons and two holes,  is dependent on the degree of its 
confinement \cite{ron12bi,eute97bi} and
the electron-to-hole ratio $\sigma$ = $\frac{m_e}{m_h}$ 
of effective masses \cite{oh1999geometric,birk1996bind}. 
Confinement effects play an important 
role in strengthening the
 biexciton binding energy \cite{ron12bi}.
The ratio of biexcitonic
$E_{biX}$ to excitonic $E_{bX}$ binding energies, $\gamma_b=\frac{E_{biX}}{E_{bX}}$
was shown as a  monotonic function of the dimensionality $\alpha$ \cite{ron12bi}.
By optimizing the geometric configuration of a simple model structure of the
biexciton where confinement effects are neglected,
Oh et. al. \cite{oh1999geometric} obtained 
$\gamma_b \approx $ 0.23 at $\sigma \approx$ 1.
The  configuration of the biexciton assumes
a  positronium-molecule-like square structure  with $\frac{r_{hh}}{a_x^{2D}}$ = 0.82,
$\frac{r_{eh}}{a_x^{2D}}$ = 0.58 \cite{oh1999geometric}.

Numerical results show that dependence of $\gamma_b$
on the  mass fraction $\sigma$ may be ignored for 
$\sigma$ values  in the range
of $\approx 0.8 -1$ \cite{ron12bi}.
A relation linking the biexciton  energies to $\alpha$ has been
obtained for $\sigma$ = 0,1 as \cite{ron12bi}
\be
\label{biener}
E_{biex}(\alpha) = \sum_{i=1}^5 \; \frac{c_i}{\alpha^{i-1}} \exp(-\alpha)
\ee
where the coefficients $c_i$ which varies according to  $\alpha$ are provided in Ref.\cite{ron12bi}.
We use a definition for the biexciton binding energy based on the energy
released when a biexciton is formed from two excitons
\be
\label{bibind}
E_{biX}(\alpha) = 2 E_{ex}(\alpha) - E_{biex}(\alpha')
\ee
where $ E_{ex}(\alpha')$ is the dimensionality dependent single exciton energy.
A negative $E_{biX}$ indicates that the biexciton is stable and
is not likely to  dissociate into two separate excitons.
It is important to note that a pair of
 excitons in two-dimensional space may be not necessarily   bind to form a biexciton
with strictly two-dimensional attributes (i.e. $\alpha \neq \alpha'$). 
We use   Eqs.\ref{biener} and \ref{bibind} to estimate the biexciton
binding energy 
\be
\label{biexe}
E_{biX} = 0.2 E_{bX} \quad \quad  \sigma \approx 1
\ee
which yields a ratio that is consistent with 
$\gamma_b \approx $ 0.23  obtained by Oh et. al. \cite{oh1999geometric}.
The pseudo two-dimensional biexciton 
has a   much larger binding energy  compared to the
 bulk biexciton. An earlier work showed that 
the  biexciton binding energy increases 
more rapidly than the excitonic binding energy when confinement is
enhanced \cite{eut}. In the case of the monolayer
 MoS$_2$, Mai et. al. \cite{mai2013many} obtained an experimental estimate of 70 meV,
and numerical binding energies in the range of 35 - 140 meV.
These values agree reasonably well with our estimated
range of 40 - 80 mev  (Eq. \ref{biexe}) for the biexciton binding
energy, by considering  the exciton binding energy  to be in the
range, 0.2 - 0.4 eV (see Table-I). The biexciton binding energies for other  Mo and 
W-family dichalcogenides  are presented in Table-II. It can be seen that the selenides
have higher static dielectric values compared to the 
sulfides, but with comparable exciton effective masses.
Consequently  the exciton, trion and biexciton binding energies
of monolayer MoS$_2$ are  higher than those of  MoSe$_2$.
The larger exciton effective mass of MoS$_2$ also results
in higher binding binding energies than those
of  WS$_2$. 

The binding energies of the biexciton and trion systems in Table-I and Table-II
 show the close spectral proximity of the exciton complexes
to the exciton quasiparticle for  various configurations of the
LTMDCs. In particular, the binding energies of 
the trion and biexciton complex quasiparticle for the monolayer configuration
 suggest favorable conditions for the  existence of the exciton-trion-biexciton states as an entangled tripartite state.
The specific features of this tripartite state is beyond the scope of this study,
and will be examined in future investigations.

\begin{table*}
    \caption{\label{tab:gw} Binding energies of the exciton, negatively charged exciton (trion)
and biexciton in select monolayer LTMDCs. Estimates for the trion and biexciton binding energies
are obtained using Eq. \ref{tri2} and \ref{biexe} respectively.
The electron ($m_e$) and hole ($m_h$) masses are retrieved from Shi et. al.\cite{shi2013quasi}
 and the corresponding exciton reduced mass, $\mu$ is compared those 
calculated using DFT+RPA by Berkelbach et. al.\cite{berkel13} (appearing within brackets).
$\epsilon$ is computed using $\epsilon_\bot,\epsilon_\|$  obtained from Kumar et. al. \cite{kumar2012tunable}.
\newline
$^\dagger$ Experimental binding energy in Ross et. al. \cite{ross2013}.
\newline
$^\ddagger$ Experimental binding energy in Jones et. al. \cite{jones2013}.
\newline
$^a$ Experimental binding energy in Komsa et. al. \cite{komsa2012effects}.
\newline
}    
\begin{tabular}{|c|c|c|c|c|c|c|c|c|c|c|c}
        \hline
        \hline
System		 ~ & $m_e$, $m_h$ &$\mu$ & $\sigma$& $\epsilon_\bot,\epsilon_\|,\epsilon$  & $E_{bX}^c$ (13.6 $\mu/\epsilon^2$) &  $E_{bX^-}$ (Eq. \ref{tri2}) & $E_{biX}$ (Eq. \ref{biexe}) \\
        \hline
         WS$_2$ &0.24, 0.31 & 0.14 (0.16) & 0.77  & 4.4, 2.9, 3.57 & 0.14 & 0.014 & 0.03 \\ 
 MoSe$_2$ & 0.36, 0.42 & 0.19 (0.27) & 0.86  & 6.9, 3.8, 5.12 & 0.10, 0.78$^a$ & 0.01, 0.03$^\dagger$ & 0.02\\
      WSe$_2$ & 0.26, 0.33 & 0.15 (0.17) & 0.79 & 4.5, 2.9, 3.61  & 0.15 & 0.015, 0.03$^\ddagger$  & 0.03 \\
        \hline
        \hline
    \end{tabular}
\end{table*}

\section{Conclusion}

In summary, we have examined the binding energies of the exciton
 in two-dimensional and  quasi-two dimensional  structures using the fractional dimensional space
approach.   The dimension of the exciton
in  LTMDCs with different number of layers is considered 
in fractional space, with the dimensionality parameter $\alpha$ seen to vary between  the ideal 2D and 3D limits.
The binding energy of exciton is linked to $\alpha$, which underlies the confinement effects of 
various geometrical structures.
The exciton binding energy decreases with increasing number
of layers in MoS$_2$, a trend that is noted in other Mo and W family of dichalcogenides.
By setting $\alpha$ = 3 in bulk  MoS$_2$, we obtain exciton binding energies
(0.2 - 0.3 eV) ($\alpha \approx$ 2) in the monolayer configuration which is 
lower than  estimates of this quantity reported in earlier works.
This energy range is consistent with experimental 
 binding energies of exciton complexes (negatively charged
trion, biexciton), as well as estimates based on known analytical relations.
We attribute a likely  reason for the disparities
in binding energies for the monolayer configuration reported in previous studies
to the underestimation of the
 dimensionality parameter $\alpha$ in  MoS$_2$. 
The main result of this study is the demonstration of the
gradual increase in $\alpha$ from 2 in monolayers, to intermediate
values (2.3 < $\alpha$ < 2.6)  in bilayer and 4-layer  MoS$_2$ system to the expected value of 
$\alpha$ = 3 in bulk  MoS$_2$. 

The application of the FDSA to LTMDCs is expected to provide  insight to critical 
opto-electronic properties which are altered via adjustment of the number of lattice layers.
The simplification provided by the fractional dimensional space
approach (FDSA) has expected tradeoffs 
in terms of not being able to 
compete with the accuracy of complicated and sophisticated  numerical approaches
afforded by  {\it ab initio} density functional techniques.
However the scaling feature of the FDSA which appear as a consequence of solving the
Schr\"odinger equation in noninteger dimensions,  can be used to
 track  non integer-dimensional occurrences of excitonic
attributes as  reliable features in  computational 
modeling.

The  band gap in MoS$_2$ and
other transition metal dichalcogenides can be engineered
by altering the number of layers and the dielectric
constant of the medium surrounding the layers.
The strength of electron-hole coulomb
interaction  and the  exciton binding energy
can also be enhanced by using specific geometrical structures
such as nanotubes   fabricated using transition metal dichalcogenides
\cite{visic2011optical}.  The increased dimensionalities of the exciton in 
such confined systems provide opportunities for selecting
attractive electro-optical properties  and specific visible features
in the optical spectra. 
Computational studies employing fractional dimensions
can therefore be used  to analyze 
 optical features in experimental spectra studies,
and facilitate the control of optical
properties of devices.

\end{document}